\documentclass{webofc}
\usepackage[varg]{txfonts}   % Web of Conferences font
\usepackage{lineno}
\usepackage{hyperref}
\begin{document}
\title{Measurements of quarkonia production}
%
% subtitle is optionnal
%
%%%\subtitle{Do you have a subtitle?\\ If so, write it here}
%\linenumbers

\author{\firstname{Laure} \lastname{Massacrier}\inst{1}\fnsep\thanks{\email{laure.marie.massacrier@cern.ch}}}

\institute{Universit\'e Paris-Saclay, CNRS/IN2P3, IJCLab, Orsay, France}

\abstract{%
  Quarkonium production in high-energy hadronic collisions is a useful tool to investigate fundamental aspects of Quantum Chromodynamics, from the proton and nucleus structure to deconfinement and the properties of the Quark Gluon Plasma (QGP). In these proceedings, emphasis is made on few recent quarkonium results from the RHIC and LHC colliders in proton-proton (pp), proton-nucleus (p-A) and nucleus-nucleus (A-A) collisions. In addition, results for some key observables are compiled to discuss the state-of-the-art in quarkonium production, with a focus on quarkonium hadroproduction. 
}
\maketitle
\section{Introduction}
\label{intro}
Quarkonium, a bound state made of a heavy quark and its antiquark, is an important tool to study Quantum Chromodynamics in high energy hadronic collisions \cite{Andronic}. Elementary pp (or e$^{+}$e$^{-}$) collisions are used to investigate quarkonium properties (production mechanisms, spectroscopy, rare decays..) and provide reference for quarkonium studies in heavy-ion collisions. Hadronic quarkonium production involves hard scales (the creation of the heavy quark-antiquark pair via hard scattering) and the subsequent soft-scale process (the binding of the pair into a colorless final state). Therefore, the quarkonium production mechanism is sensitive to the gluon content of the colliding proton and thus provides information on parton distribution functions (PDFs) \cite{Ozcelik}. In p-A collisions, the influence of a nucleus in the initial state on quarkonium production can be investigated. These effects, known as Cold Nuclear Matter effects (CNM), include among others: the modification of parton densities in nuclei (nPDFs) \cite{Eskola}, coherent energy-loss effects involving the initial- and final-state partons \cite{Arleo}, nuclear absorption \cite{Ferreiro1} and the interaction of the quarkonium state with other particles produced in the collision (the comovers) \cite{Ferreiro2}.  In A-A collisions, quarkonium is a sensitive probe of the formed hot strongly interacting medium, as hard-scattering processes occur at the early stage of the collision, on a timescale that is in general shorter than the QGP lifetime. Quarkonium production is expected to be suppressed in the QGP due to static color screening \cite{Matsui} or dynamical dissociation \cite{Rothkopf}, the suppression being sequential, according to the binding of each quarkonium state. In this simple picture, the in-medium dissociation of the quarkonium state should provide an estimate of the initial temperature reached in the collision. However, at high energy, if the number of heavy quark pairs produced is large enough, quarkonium recombination \cite{Braun,Thews} can take place. Evidence for such phenomenon exist at LHC energies for charmonium \cite{ALICE1}, and it directly points to the existence of a deconfined QGP. It is worth mentioning that low transverse momentum ($p_{\rm T}$) heavy quarks could participate, through their interactions with the medium, in the collective expansion of the system and possibly reach thermal equilibrium with its constituent. Recently, several measurements in high multiplicity pp and p-Pb collisions at LHC, have revealed the presence of phenomena typically attributed to the creation of a QGP (eg. first observation of ridge in two-particle azimuthal correlations \cite{CMS1}). Multiplicity-dependent quarkonium studies offer a valuable testing ground for examining the onset of QGP-like effects in small systems, while also probing the interplay between soft and hard particle production and the role of multiparton interactions (MPI). 

%sentence for high multiplicity pp/pA quarkonium production 
\section{Quarkonium production in proton-proton collisions}

\begin{figure}[!htpb]
\begin{minipage}{0.5\linewidth}
\centerline{\includegraphics[width=1.\linewidth]{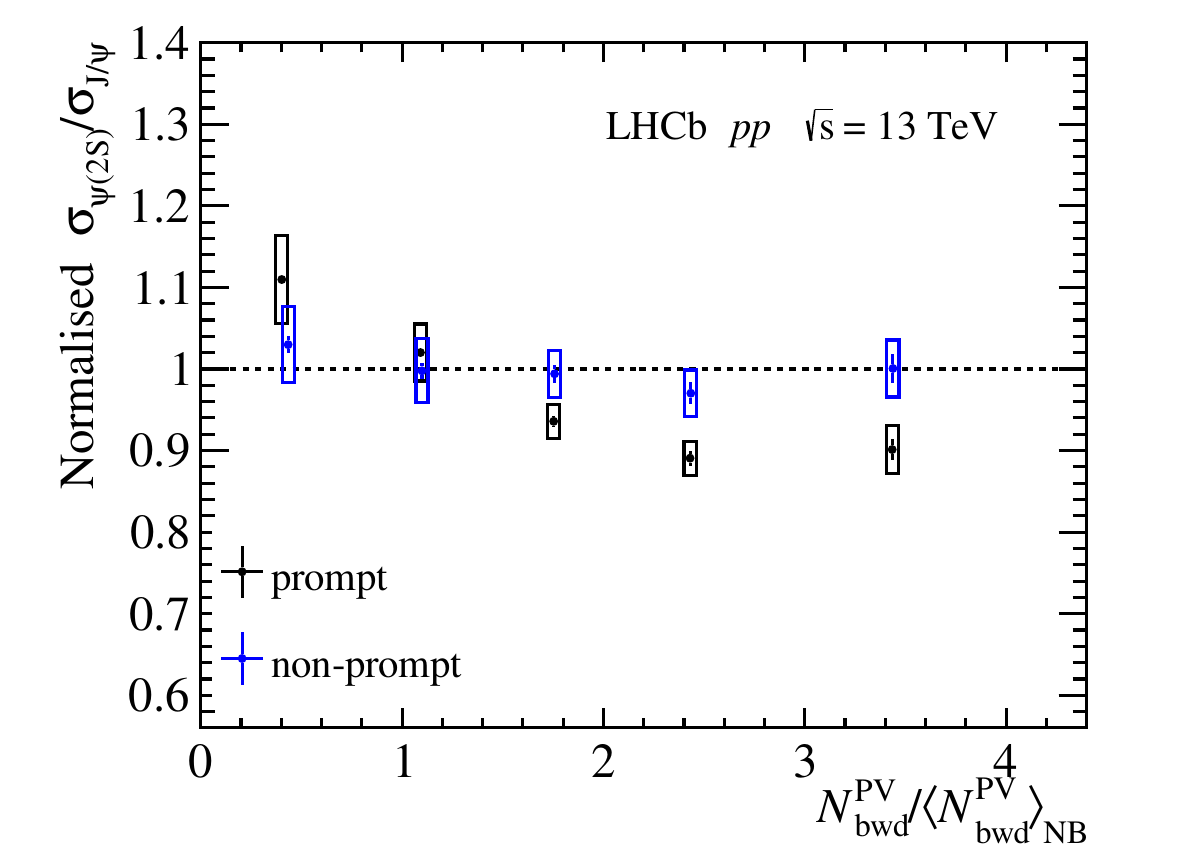}}
\end{minipage}
\hfill
\begin{minipage}{0.5\linewidth}
\centerline{\includegraphics[width=1.\linewidth]{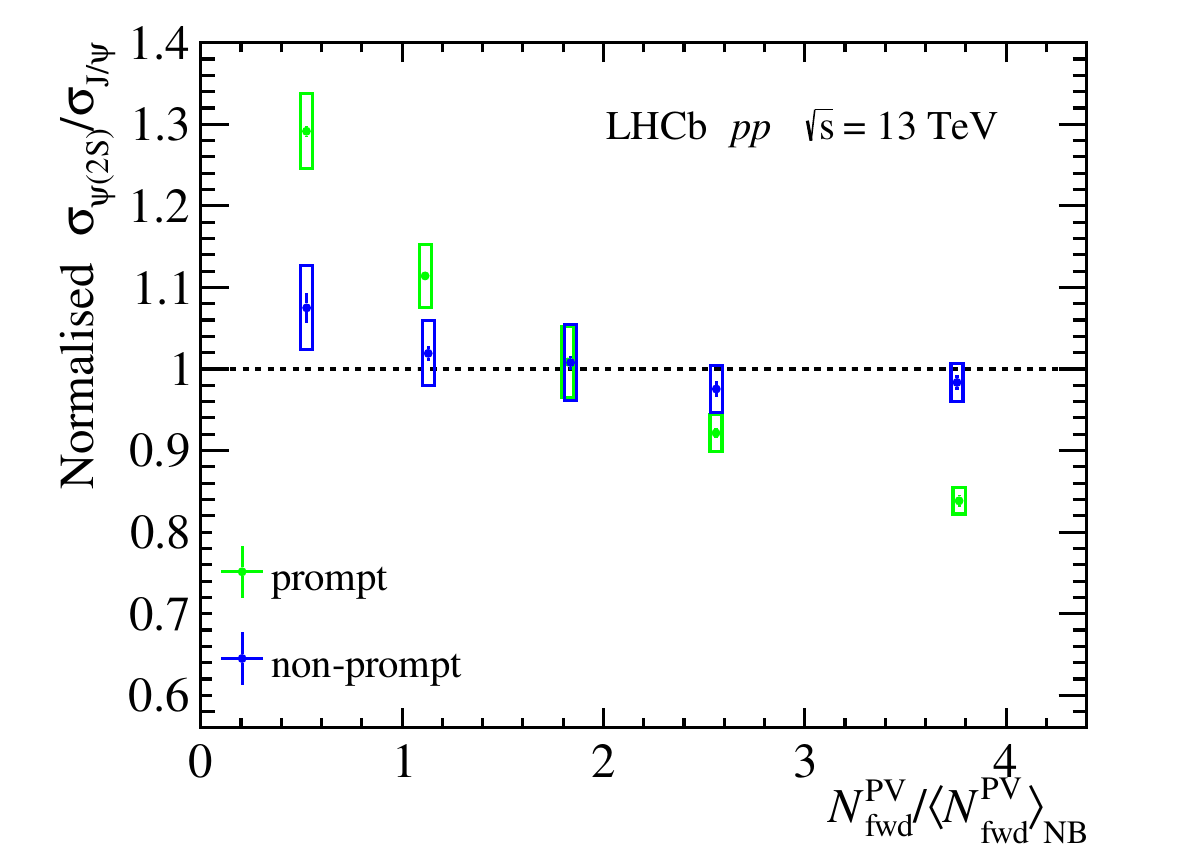}}
\end{minipage}
\hfill
\caption[]{Self-normalized $\psi$(2S)-to-J/$\psi$ cross section ratio as a function of the number of tracks (considered for primary vertex reconstruction) measured by the VELO of LHCb, in the backward (left) and forward (right) pseudorapidity regions, in pp collisions at $\sqrt{s}$ = 13 TeV. Results for (non-prompt) prompt charmonium are reported in (blue) black (left) and (blue) green (right) \cite{LHCb1}.}
\label{fig1}
\end{figure}

Prompt quarkonium production is in general rather well described over a wide range of rapidity, $p_{\rm T}$ and center-of-mass (c.o.m) energy, by Non Relativistic QCD (NRQCD) \cite{NRQCD} and Improved Color Evaporation (ICEM) \cite{Cheung} models. In recent years, LHC quarkonium data are becoming more precise than theoretical model uncertainties. In order to further constrain models and therefore quarkonium production mechanisms, the ALICE Collaboration has measured the ratio of J/$\psi$ and $\psi$(2S) cross sections at various c.o.m energies to the ones obtained at $\sqrt{s}$ = 13 TeV \cite{ALICE2}. These ratios benefit from the partial cancellation of several theoretical scale uncertainties and it was shown that it becomes challenging for models to reproduce quarkonium cross section ratios. The simultaneous description of quarkonium production cross section and polarization has also been a long standing puzzle for models. Indeed, S-wave quarkonia are produced almost unpolarized at LHC \cite{Andronic}, challenging NRQCD predictions favoring sizable polarization, especially at high $p_{\rm T}$. However, recent results from CMS on $\chi_{c1}$ and $\chi_{c2}$ P-wave quarkonia \cite{CMS2} favor a scenario where at least one of the two states is strongly polarized, as expected from NRQCD. Recent developments in NRQCD using global fit analyses were able to improve the description of quarkonium S-wave polarization results \cite{Faccioli}. Studies at both RHIC and LHC have revealed that quarkonium production shows a noticeable dependence on the event multiplicity \cite{ALICE3,PHENIX1}, with increasing yields as multiplicity rises. This suggests that quarkonium production is influenced by the overall activity in the event, hinting at complex interactions between hard and soft processes. The enhanced quarkonium production rate with multiplicity is usually interpreted as a result of MPI. Excited to ground state quarkonium yield ratio as a function of the charged particle multiplicity can permit to further pin down final state effects on quarkonia production. In the comover scenario \cite{Ferreiro2}, quarkonia can be dissociated by interaction with the comoving particles in the final state, with the dissociation probability depending on the binding energy of each quarkonium state and on the density of comoving particles. Figure~\ref{fig1} shows recent measurements from LHCb on the self-normalized $\psi$(2S)-to-J/$\psi$ cross section ratio as a function of the number of tracks considered for primary vertex reconstruction in the backward (left) and forward (right) pseudorapidity regions \cite{LHCb1}. Results are shown separately for prompt and non-prompt charmonia. While the ratio of charmonia from b-hadrons shows no evident multiplicity dependence irrespective of the region in which the multiplicity is measured, the prompt $\psi$(2S)-to-J/$\psi$ ratio exhibits a decreasing trend with multiplicity for both multiplicity estimators. The prompt $\psi$(2S) are more suppressed than the prompt J/$\psi$ for high forward multiplicities\footnote{Note that the muon tracks from the charmonium decays are included in the forward multiplicity estimator} (i.e  when the multiplicity is measured in a similar acceptance as the charmonia), in line with expectations from a comover scenario. This suggests a correlation between the observed suppression and the local particle multiplicity. On the other hand, the fact that the $\psi$(2S)-to-J/$\psi$ ratio is not flat as a function of the backward multiplicity, as expected in a naive comover picture, could be attributed to the existence of strong correlations between the forward and backward multiplicities (see Fig.~4 of \cite{LHCb1}). These studies demonstrate the importance of performing a systematic set of studies varying the quarkonium and multiplicity rapidity ($y$) windows, for proper interpretation of quarkonium versus multiplicity results.

\section{Quarkonium production in proton-nucleus collisions}

\begin{figure}[!htpb]
\begin{minipage}{0.5\linewidth}
\centerline{\includegraphics[width=0.9\linewidth]{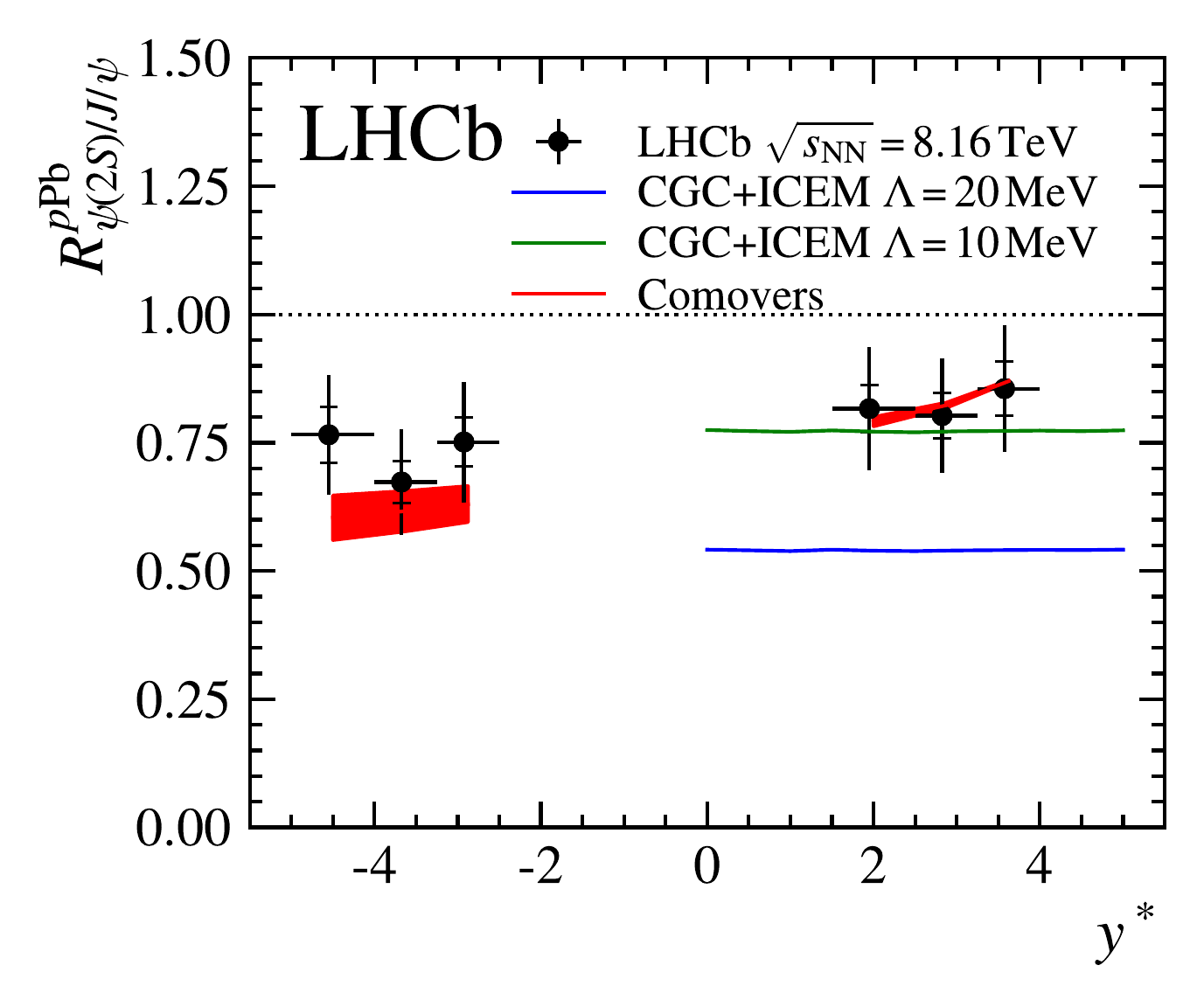}}
\end{minipage}
\hfill
\begin{minipage}{0.5\linewidth}
\centerline{\includegraphics[width=1.\linewidth]{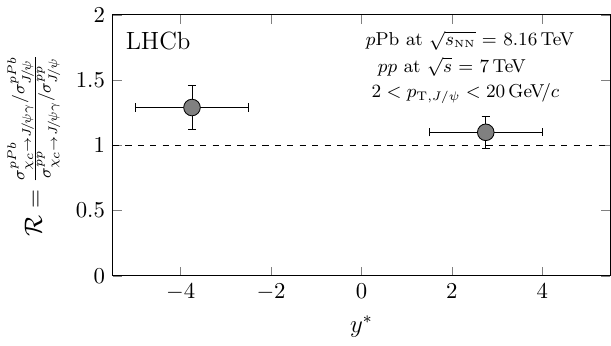}}
\end{minipage}
\hfill
\caption[]{Left: R$^{\rm pPb}_{\rm \psi(2S)/J/\psi}$ for prompt charmonium, as a function of the c.o.m rapidity $y^{\star}$ measured by LHCb in p--Pb collisions at $\sqrt{s_{\rm NN}}$ = 8.16 TeV \cite{LHCb3}. Data are compared to a CGC soft gluon interaction model \cite{Ma} and a comover model \cite{Ferreiro2}. Right: Double ratio $\sigma_{\chi_{c} \rightarrow J/\psi \gamma}$/$\sigma_{J/\psi}$ between p--Pb collisions at $\sqrt{s_{\rm NN}}$~=~8.16~TeV and pp collisions at $\sqrt{s}$~=~7~TeV, as a function of $y^{\star}$, measured by LHCb \cite{LHCb5,LHCb6}. The measured $\chi_{c}$ cross section is the sum of  $\chi_{\rm c1}$ and  $\chi_{\rm c2}$ cross sections.}
\label{fig2}
\end{figure}

\begin{figure}[!htpb]
\begin{minipage}{0.5\linewidth}
\centerline{\includegraphics[width=0.8\linewidth]{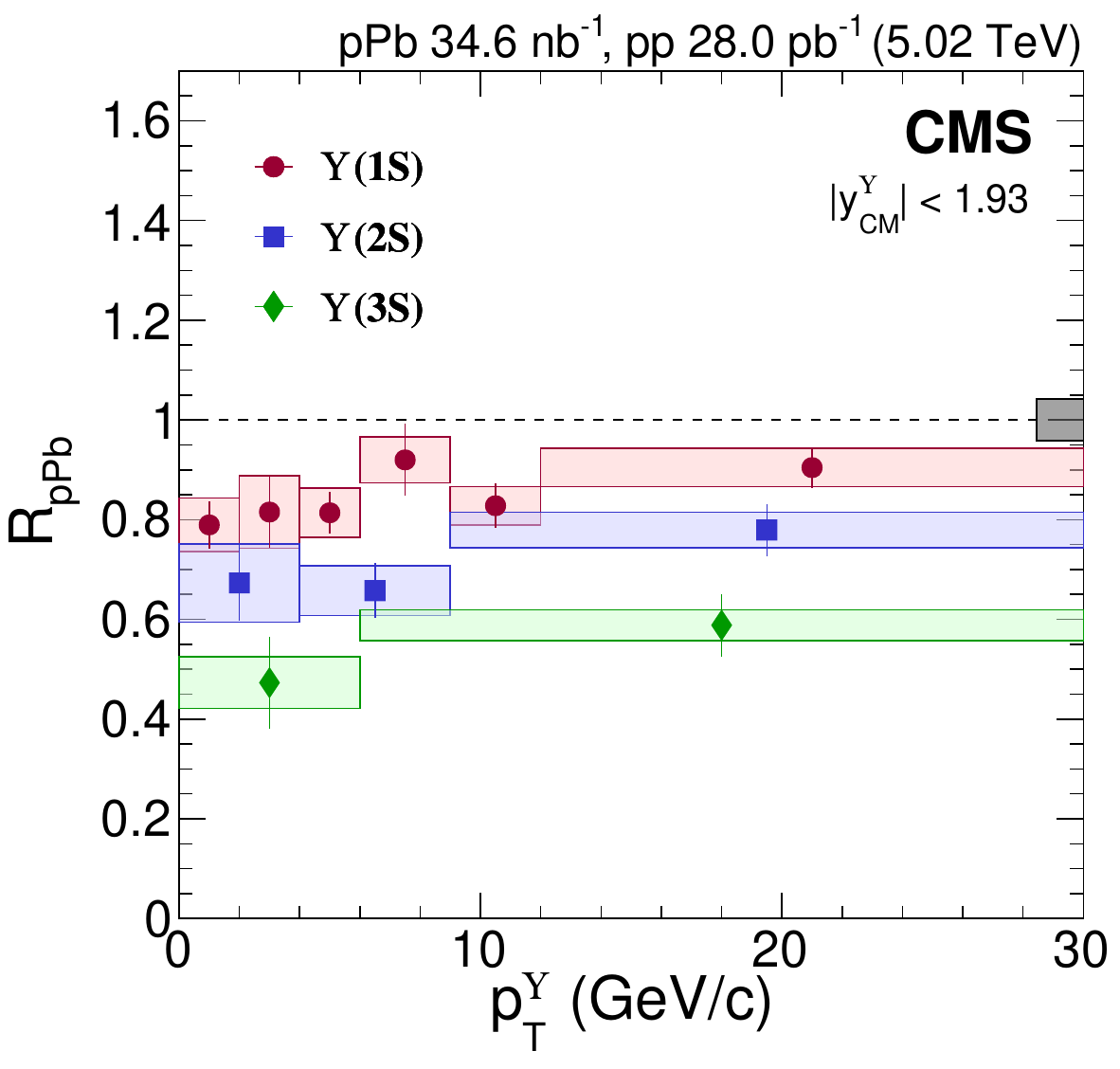}}
\end{minipage}
\hfill
\begin{minipage}{0.5\linewidth}
\centerline{\includegraphics[width=1.1\linewidth]{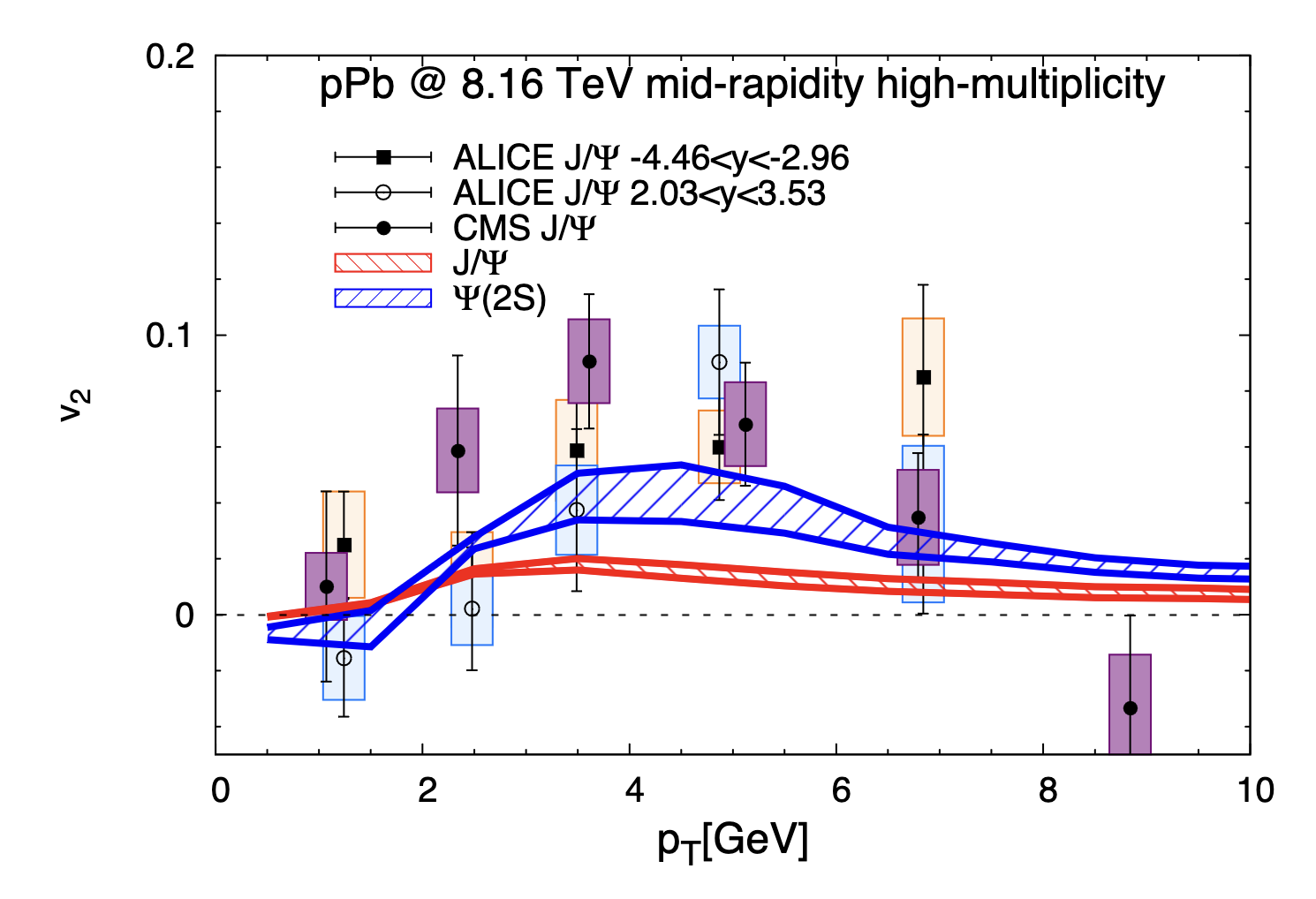}}
\end{minipage}
\hfill
\caption[]{Left: R$_{\rm pPb}$ of $\Upsilon{\rm (1S)}$ (red), $\Upsilon{\rm (2S)}$ (blue), $\Upsilon{\rm (3S)}$ (green) as a function of $p_{\rm T}$ measured at midrapidity by CMS in p--Pb collisions as $\sqrt{s_{\rm NN}}$ = 5.02 TeV \cite{CMS3}. Right: Transverse-momentum dependent elliptic flow of J/$\psi$ in high multiplicity p--Pb collisions at $\sqrt{s_{\rm NN}}$ = 8.16 TeV, measured by ALICE at forward (open circle) and backward (full square) rapidities \cite{ALICE6}, and by CMS at midrapidity (full circle) \cite{CMS4}. Data are compared to a transport model from \cite{Du}.}
\label{fig3}
\end{figure}

The production of the J/$\psi$ charmonium in p--A collisions has been extensively studied from RHIC to the LHC energies \cite{Andronic}. The J/$\psi$ nuclear modification factor (R$_{\rm pPb}$) at LHC energies exhibits almost no suppression at backward rapidity \cite{ALICE4,LHCb2}, moderate suppression at midrapidity mainly occurring in the low-$p_{\rm T}$ region \cite{ALICE5} and stronger suppression at forward rapidity \cite{ALICE4,LHCb2}. This behavior is in general well reproduced by several nPDFs sets (see reference therein), and coherent energy loss model \cite{Arleo}. These calculations predict similar R$_{\rm pPb}$ for J/$\psi$ and $\psi$(2S) as they assume that the nuclear suppression is prior to the hadronisation timescale, and that the nuclear effects apply similarly to both resonances given their proximity in mass. Figure~\ref{fig2} left shows recent LHCb measurement of the prompt $\psi$(2S) R$_{\rm pPb}$ to J/$\psi$ R$_{\rm pPb}$ ratio as a function of the c.o.m rapidity, in p--Pb collisions at $\sqrt{s_{\rm NN}}$ = 8.16 TeV \cite{LHCb3}. The results indicate a stronger suppression for the prompt $\psi$(2S) with respect to the prompt J/$\psi$ in all $y$ intervals, and slightly more pronounced at backward rapidity. The observed behaviour can be described by models incorporating final state interactions breaking preferentially the $\psi$(2S) meson compared to the J/$\psi$ meson \cite{Ferreiro2}. However, recent measurements of the excited $\chi_{c}$ charmonium states in p--Pb collisions could challenge these interpretations. Figure~\ref{fig2} right illustrates the $\chi_{c}$ to prompt J/$\psi$ cross section ratio between p--Pb and pp collisions, measured by LHCb \cite{LHCb5,LHCb6}. The ratio is close to unity at both backward and forward rapidities, indicating no significant cold nuclear matter effects on $\chi_{c}$, despite the similar binding energies of $\psi$(2S) and $\chi_{c}$ states. In the bottomonia sector, recent results from CMS at midrapidity on the $\Upsilon$(nS) R$_{\rm pPb}$ as a function of $p_{\rm T}$ are shown in Fig.~\ref{fig3} (left) \cite{CMS3}. The $\Upsilon$(nS) states follow an ordered suppression pattern, similar to the one observed in the charmonium sector for $\psi$(2S) with respect to J/$\psi$. Indeed, the measured R$_{\rm pPb}$($\Upsilon$(1S)) is systematically larger than that of $\Upsilon$(2S), which in turn is systematically larger than the R$_{\rm pPb}$($\Upsilon$(3S)), also suggesting different level of modifications for the three states by final state effects. Measurement of quarkonium elliptic flow (v$_{2}$) in high-multiplicity p--Pb collisions are complementary to the ones for light flavor particles, in order to understand the collective behavior of particles in small collision systems. Models based on the hydrodynamic expansion of a tiny QGP droplet, or alternative scenarios based on gluon saturation in the initial state have been developed in order to interpret the long-range collective azimuthal correlations observed in high multiplicity pp and p--Pb collisions \cite{Dusling}. Figure~\ref{fig3} right shows the transverse-momentum dependent elliptic flow of J/$\psi$ in high multiplicity p--Pb collisions measured by ALICE \cite{ALICE6} and CMS \cite{CMS4} in different rapidity windows. Positive J/$\psi$ v$_{2}$ values are measured in both experiments for about $p_{\rm T} >$~3~GeV/c, indicating charm quark collectivity in high multiplicity p--Pb events. The collective effects are found to be weaker for charm quarks than for light quarks \cite{CMS5}. Data at midrapidity are compared with a transport model calculation including path-length effects for primordial charmonium traversing the elliptic fireball, and a non-zero v$_{2}$ for regenerated charmonia \cite{Du}. The model strongly underestimates the J/$\psi$ v$_{2}$, suggesting that the observed v$_{2}$ in data cannot come from final-state interactions alone. The fact that ALICE measures similar J/$\psi$ v$_{2}$ at backward and forward rapidities, while the reached multiplicity is rather different in both rapidity regions, is also in line with this conclusion.

\section{Quarkonium production in nucleus-nucleus collisions}

\begin{figure}[!htpb]
\begin{minipage}{0.5\linewidth}
\centerline{\includegraphics[width=1.05\linewidth]{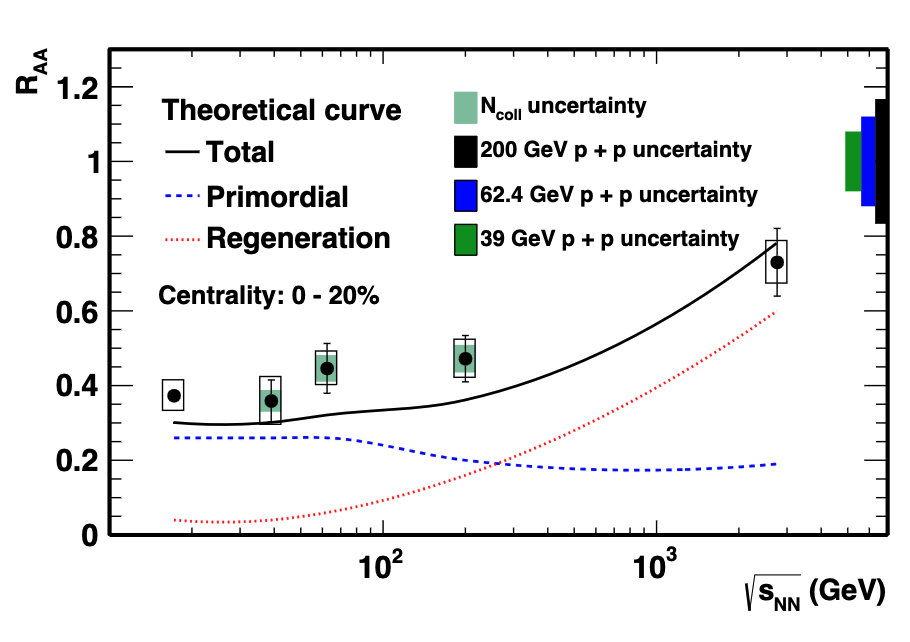}}
\end{minipage}
\hfill
\begin{minipage}{0.5\linewidth}
\centerline{\includegraphics[width=0.95\linewidth]{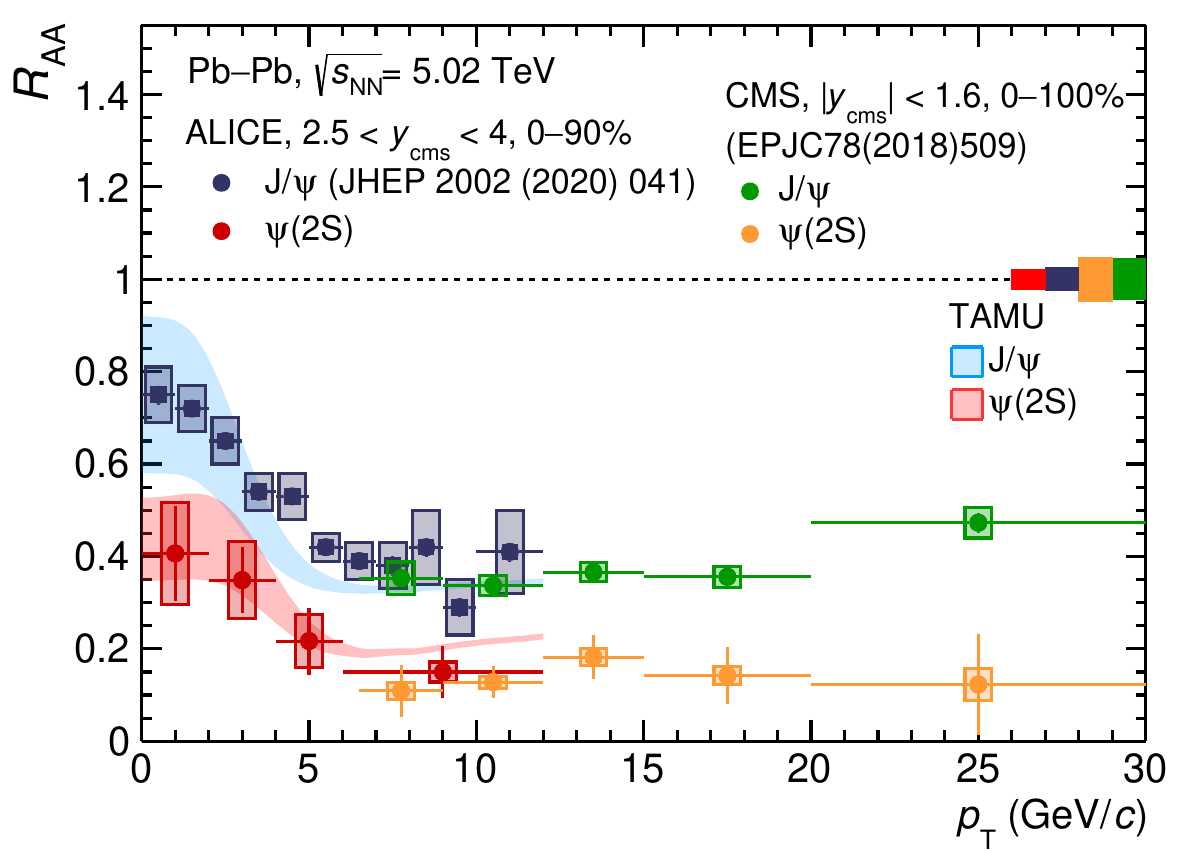}}
\end{minipage}
\hfill
\caption[]{Left: J/$\psi$ R$_{\rm AA}$ as a function of $\sqrt{s_{\rm NN}}$ measured at SPS \cite{SPS1}, RHIC \cite{STAR1} and the LHC \cite{ALICE7} in most central A--A collisions. Data are compared to the transport model calculations from \cite{Zhao}.  Right: J/$\psi$ and $\psi$(2S) R$_{\rm AA}$ as a function of $p_{\rm T}$ measured by ALICE at forward rapidity \cite{ALICE8,ALICE9} and CMS at midrapidity \cite{CMS6}, in Pb--Pb collisions at $\sqrt{s_{\rm NN}}$ = 5.02 TeV. Results are compared to the transport model from \cite{Du2}.}
\label{fig4}
\end{figure}

\begin{figure}[!htpb]
\begin{minipage}{0.5\linewidth}
\centerline{\includegraphics[width=1\linewidth]{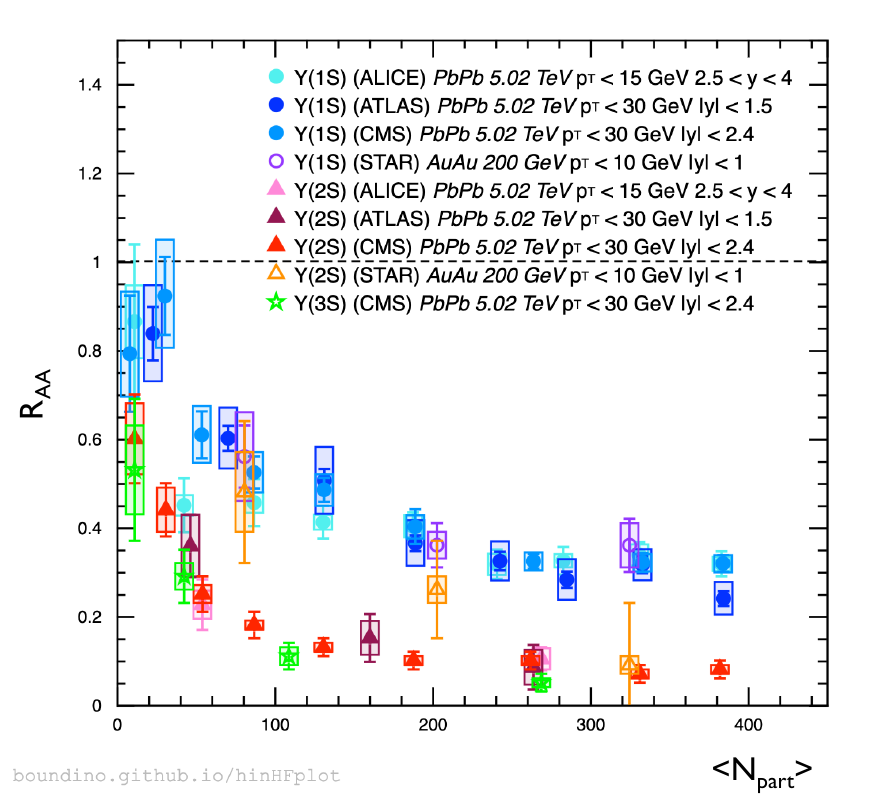}}
\end{minipage}
\hfill
\begin{minipage}{0.5\linewidth}
\centerline{\includegraphics[width=1.\linewidth]{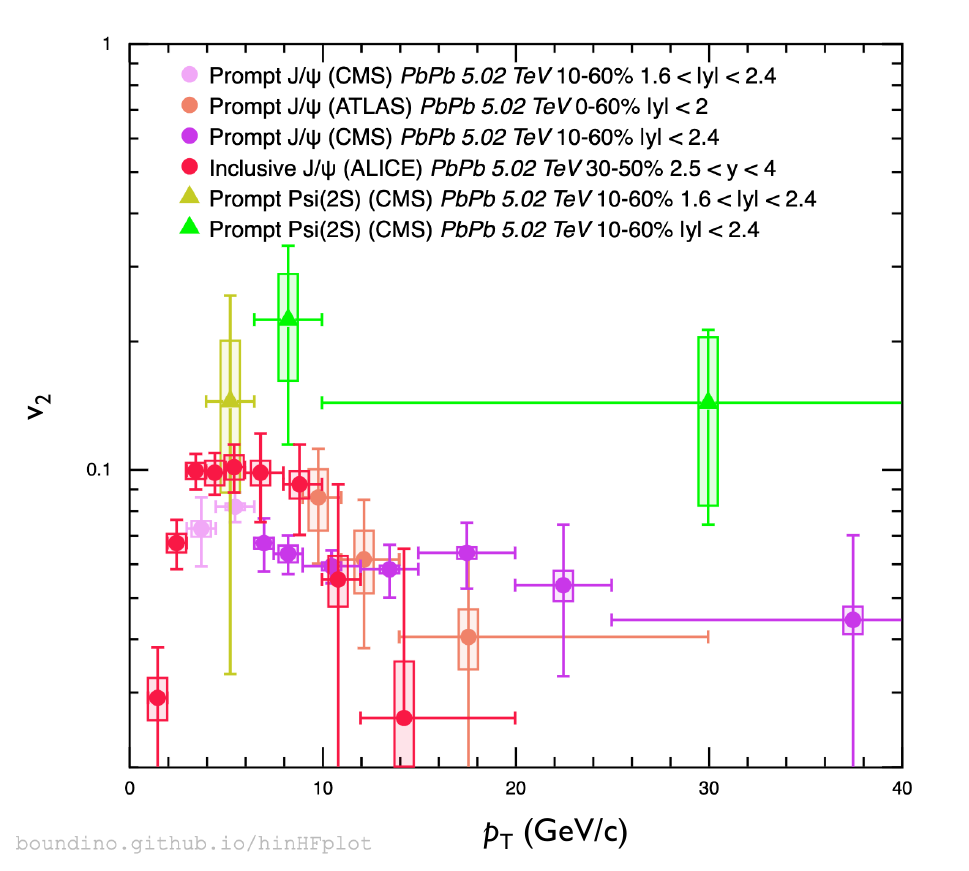}}
\end{minipage}
\hfill
\caption[]{Left: $\Upsilon$(nS) R$_{\rm AA}$ as a function of the mean number of participant nucleons <N$_{\rm part}$>, measured at RHIC \cite{STAR1} in Au--Au collisions and the LHC \cite{CMS7,CMS8,ALICE10,ATLAS1} in Pb--Pb collisions. Right: Charmonium elliptic flow as a function of $p_{\rm T}$ measured by ALICE \cite{ALICE11}, ATLAS \cite{ATLAS2} and CMS \cite{CMS9} in semi-central Pb--Pb collisions at $\sqrt{s_{\rm NN}}$ = 5.02 TeV. These compilation plots are generated from \cite{HFplots}.}
\label{fig5}
\end{figure}

Quarkonia are long standing probes of the QGP. In a hot and deconfined medium, they are suppressed by static color screening \cite{Matsui} or dynamical dissociation \cite{Rothkopf}. This suppression should occur sequentially, according to the binding energy of each mesons: strongly bound states such as $\Upsilon$(1S) and J/$\psi$ are expected to melt at higher temperatures with respect to the loosely bound states (e.g. $\chi_{b}$, $\Upsilon$(2S), $\Upsilon$(3S) for the bottomonia family, $\psi$(2S) and $\chi_{c}$ for the charmonia family). However, the prediction of a sequential suppression pattern is complicated by several factors, such as feed-down (from higher mass resonances, or b-hadrons in the case of charmonium), as well as other hot nuclear matter effects (such as recombination), or cold nuclear matter effects. A comprehensive set of measurements of several quarkonium states is therefore crucial to interpret QGP effects on quarkonium production. Figure~\ref{fig4} left shows the J/$\psi$ R$_{\rm AA}$ measured at various c.o.m energies, from SPS \cite{SPS1} and RHIC \cite{STAR1} to the LHC \cite{ALICE7}, in most central A--A collisions (Pb--Pb for SPS/LHC and Au--Au for RHIC). No significant energy dependence of the J/$\psi$ R$_{\rm AA}$ is observed for $\sqrt{s_{\rm NN}} <$ 200 GeV, while an important increase is seen at LHC energies. Data are compared to the transport model from Ref. \cite{Zhao} which includes two main components. The primordial component represents the anomalous suppression of primordial J/$\psi$ due to CNM and color screening effects. If only CNM effects are considered, the R$_{\rm AA}$ is expected to be around 0.6. The regeneration component arises from the combination of uncorrelated $c\bar{c}$ pairs in medium. In this model, the feed-down from $\psi$(2S) and $\chi_{c}$ to the J/$\psi$ are accounted for. As the QGP temperature increases, the J/$\psi$ suppression by color screening becomes more important. At the same time, the regeneration component increases with collision energy due to the increased production of charm quark pairs, leading to a compensation of the stronger suppression. The higher J/$\psi$ R$_{\rm AA}$ at LHC tends to indicate that the surviving J/$\psi$ are mainly originating from a recombination contribution. In order to better constrain regeneration models, the comparison of the $\psi$(2S) and J/$\psi$ yield modifications in A--A collisions is a golden probe. Figure~\ref{fig4} right shows the J/$\psi$ and $\psi$(2S) R$_{\rm AA}$ measured by ALICE at forward rapidity \cite{ALICE8,ALICE9} and CMS at midrapidity \cite{CMS6} in Pb--Pb collisions. At high $p_{\rm T}$, both $\psi$(2S) and J/$\psi$ are strongly suppressed, the suppression being the strongest for the $\psi$(2S) state. The $\psi$(2S) R$_{\rm AA}$ hints at an increase towards low $p_{\rm T}$, following a similar pattern as for the J/$\psi$, which is understood as a direct consequence of the regeneration process for charm and anticharm quarks. The J/$\psi$ and $\psi$(2S) R$_{\rm AA}$ are well described as a function of $p_{\rm T}$ by the transport model from Ref. \cite{Du2} which includes charmonium regeneration in the QGP phase. The bottomonium family is expected to provide additional information on the thermal properties of the deconfined medium. The $\Upsilon$(1S) resonance is the most tightly bound quarkonium state and therefore has one of the highest dissociation temperature ($\sim$ 2T$_{\rm c}$, with T$_{\rm c}$ the critical temperature for deconfinement). In addition, recombination-like processes for bottomonia are expected to be negligible compared to charmonia at LHC energies, as the number of beauty quark pairs per event is much smaller compared to the number of charm quark pairs. Figure~\ref{fig5} left shows the $\Upsilon$(nS) R$_{\rm AA}$ as a function of <N$_{\rm part}$>, measured at RHIC \cite{STAR1} and the LHC \cite{CMS7,CMS8,ALICE10,ATLAS1}. A gradual decrease of the $\Upsilon$(nS) R$_{\rm AA}$ is observed with increasing <N$_{\rm part}$> at LHC energies. The $\Upsilon$(3S) resonance is measured for the first time in Pb--Pb collisions by CMS \cite{CMS7} and is found to be strongly suppressed in central Pb--Pb collisions. A sequential suppression pattern of the $\Upsilon$(nS) states is observed at LHC energy, and indication for such pattern is also visible at RHIC in central Au--Au collisions for the $\Upsilon$(1S) and $\Upsilon$(2S) states, although with a smaller significance. Remarkably, the $\Upsilon$(1S) R$_{\rm AA}$ gets similar values at LHC and RHIC energies, despite a factor $\sim$ 25 difference in $\sqrt{s_{\rm NN}}$. This observation is in favor of a negligible melting of the direct $\Upsilon$(1S) state at the probed energies, but might be explained by the suppression of the excited states in combination with CNM effects. Being sensitive to the dynamics of the partonic stages of heavy-ion collisions, the v$_{2}$ observable can provide complementary information to the R$_{\rm AA}$, in particular on the quarkonium production mechanisms and the degree of thermalization of heavy quarks in medium. Indeed, charmonia produced via regeneration should inherit the elliptic flow of charm quarks in the QGP. On the other hand, primordial charmonia can acquire v$_{2}$ because of path-length dependent suppression through the medium. This latter contribution becomes dominant at high $p_{\rm T}$, while the influence of a regeneration component should dominate at low $p_{\rm T}$. Figure~\ref{fig5} right shows a compilation of inclusive and prompt J/$\psi$ v$_{2}$ results at the LHC in semi-central Pb--Pb collisions \cite{CMS9,ALICE11,ATLAS2}, as well as the first prompt $\psi$(2S) v$_{2}$ measured by CMS \cite{CMS9}. A significant J/$\psi$ v$_{2}$ is observed at low-$p_{\rm T}$, stronger than in p--Pb collisions, favoring a scenario with significant regeneration of J/$\psi$ in Pb--Pb collisions at the LHC. Given the lower binding energy of the $\psi$(2S) state, the recombination of $\psi$(2S) can occur at a later stage of the QGP evolution, leading to a larger v$_{2}$ for the regenerated $\psi$(2S) with respect to the regenerated J/$\psi$ \cite{Du3}. The prompt $\psi$(2S) v$_{2}$ measured by CMS tends to exhibit similar, and possibly larger v$_{2}$ than the J/$\psi$, albeit large uncertainties. More precise $\psi$(2S) measurements extended towards lower $p_{\rm T}$ would help probing the $\psi$(2S) production mechanisms in A--A collisions. It is worth mentioning that bottomonia (charmonia) measurements at LHC\cite{ALICE12,CMS10} (RHIC\cite{STAR2}) in A--A collisions do not exhibit significant v$_{2}$ within the large experimental uncertainties. 

%RHIC charmonium flow compatible with zero albeit large uncertainties 

\section{Conclusion}

A large variety of precise quarkonium cross section measurements have been performed at both RHIC and the LHC, in pp collisions, and are overall rather well described by NRQCD and ICEM calculations. Recent improvements in the simultaneous description of  quarkonium production cross section and polarization have been made by the same models. Quarkonium production as a function of the event multiplicity has demonstrated to be a powerful tool to study MPI and final state interaction (including interactions with comoving particles) on quarkonium production. In pA collisions at the LHC, the systematic studies of several charmonia and bottomonia states allow for a better understanding of CNM effects (nPDFs, coherent energy loss) on quarkonium production, including the role of feed-downs. Weakly bound resonances ($\psi$(2S),$\Upsilon$(3S)) seem to be suppressed with respect to their vector meson ground states (except for $\chi_{c}$), indicating the need for final state effects to describe quarkonium production in pA. The non-zero J/$\psi$ v$_{2}$ measured in high multiplicity pA collisions enlightens heavy quark collectivity however cannot be explained by final state interactions only. In AA collisions, charmonia follow a sequential suppression pattern at RHIC, while at LHC energies there is a clear evidence for charmonium regeneration competing with the sequential dissociation. Whether charmonium regeneration takes place throughout the QGP lifetime or at the phase transition cannot be disentangled with current data, and would require for instance more precise measurement of charmonium excited states. The elliptic flow measurement of J/$\psi$ in AA collisions at LHC exhibits sign of charm quark thermalization in medium. In the bottomonia sector, $\Upsilon$(nS) states tend to exhibit a sequential suppression pattern at both RHIC and the LHC. The combination of both measurements seems to indicate that the temperature reached in the QGP is not large enough to melt the direct $\Upsilon$(1S) state, even at LHC energies.

\section*{Acknowledgments}

The author acknowledges the support of the French Agence Nationale de la Recherche (ANR) under reference ANR-22-CE31-0005 MALICE.

\end{document}